# MAXIMUM ENTROPY METHOD: SAMPLING BIAS

Jorge Fernandez-de-Cossio [ɫ] & Jorge Fernandez-de-Cossio Diaz[Ŧ]

ɫ Center for Genetic Engineering and Biotechnology, Havana Cuba, Ŧ Center for molecular Immunology, Havana, Cuba.

Corresponding author jorge.cossio@cigb.edu.cu

## Abstract

Maximum entropy method is a constructive criterion for setting up a probability distribution maximally non-committal to missing information on the basis of partial knowledge, usually stated as constrains on expectation values of some functions. In connection with experiments sample average of those functions are used as surrogate of the expectation values. We address sampling bias in maximum entropy approaches with finite data sets without forcedly equating expectation values to corresponding experimental average values. Though we rise the approach in a general formulation, the equations are unfortunately complicated. We bring simple case examples, hopping clear but sufficient illustration of the concepts.

## Significant statement

The impacting success of descriptions of diverse experiments and phenomena in terms of an inverse problem of statistical mechanics urge the attention to bias introduced by under sampling. Such is the case of large biological networks where sample collection is severely limited.  Here we address sampling bias in maximum entropy approaches with limited data sets.

## Introduction

Important technological advances in the last couple of decades have permitted the collection of huge amounts of detailed data from a wide variety of systems with large number of degrees of freedom. A recent outbreak of methods borrowed from thermodynamics, statistical physics and information theory are being applied to the analysis and interpretation of these datasets. In general, the problem can be described as a search for a connection between macroscopic behavior and the detailed interactions of many actors at the microscopic scale. Diverse systems (neuroscience, network biology, flocking, finance, sociology) have now been described as collective phenomena, where non-trivial properties even differing in their microscopic details exhibit similar thermodynamic behavior (Asta, Castellano, & Marsili, 2007; Bialek, Cavagna, Giardina, Mora, & Silvestri, 2012; Braunstein, Pagnani, Weigt, & Zecchina, 2008; de Lachapelle & Challet, 2010; Schneidman, Berry II, Segev, & Bialek, 2006).

The principle of maximum entropy dates back to a reinterpretation of statistical mechanics where thermodynamics and information entropy emerge as the same concept (Jaynes, 1957). Maximum entropy is a constructive criterion for setting up a probability distribution maximally non-committal to missing information on the basis of partial knowledge.

The principle can be stated in this way: Inference is required on the properties of a system whose state is determined by a set of parameters $\boldsymbol{\sigma}$. The available information about the system can in many instances be restated as testable constrains on the probability distribution $p(\boldsymbol{\sigma})$. If these constrains take the form of expectation values $M = \{M_1, \dots, M_m\}$ of some functions $f_1(\boldsymbol{\sigma}), \dots, f_m(\boldsymbol{\sigma})$, the probability distribution $p(\boldsymbol{\sigma})$ expressing maximum uncertainty with respect to all other unspecified matters is the one which maximizes the entropy

$$S = -\sum_{\sigma} p(\boldsymbol{\sigma}) \log p(\boldsymbol{\sigma})$$

subject to the constrains

$$\sum_{\sigma} f_j(\boldsymbol{\sigma}) p(\boldsymbol{\sigma}) = M_j, \qquad j = 1, \dots, m,$$

and $\sum_{\sigma} p(\boldsymbol{\sigma}) = 1$. The well-known solution to this problem is the Gibbs generalized canonical ensemble

$$p(\boldsymbol{\sigma}) = \frac{1}{Z} \exp\{-E(\boldsymbol{\sigma}|\lambda)\}$$

where

$$E(\boldsymbol{\sigma}|\lambda) = \lambda_1 f_1(\boldsymbol{\sigma}) + \dots + \lambda_m f_m(\boldsymbol{\sigma})$$

and the constants $\lambda_j$ and $Z$ are chosen so that the constrains are satisfied. Borrowing the terminology from statistical physics, $E(\boldsymbol{\sigma}|\lambda)$ is called the energy of the system, the normalization constant $Z$ is also called the partition function,

$$Z(\lambda_1, \dots, \lambda_m) = \sum_{\sigma \in \Sigma} \exp\{-E(\boldsymbol{\sigma}|\lambda)\},$$

and the Helmholtz free energy $F$ is defined as

$$F(\lambda_1, \dots, \lambda_m) = -\log Z(\lambda_1, \dots, \lambda_m).$$

The constants $\lambda_j$, also called Lagrange multipliers, satisfy the relations:

$$M_j = \frac{\partial F}{\partial \lambda_j}, \qquad \lambda_j = \frac{\partial S}{\partial M_j}$$

Inferring the Lagrange multipliers is in general hard and searching for a solution is usually called the inverse problem. Recent proposed procedures are basically maximum likelihood approaches, requiring additional regularization techniques to avoid over fitting the data.

In most applications of this method the expectation values $M$ are unknown. Instead, datasets of independent samples $D = \{\boldsymbol{\sigma}_1, \dots, \boldsymbol{\sigma}_n\}$ allow us to take sample averages of the functions $f_j(\sigma)$ as a proxy:

$$\overline{f_j} = \frac{1}{n} \sum_{i=1}^{n} f_j(\boldsymbol{\sigma}_i)$$

When the sample size $n$ is much larger than the dimension of the value space of constrains, the approximation $M_j \approx \overline{f_j}$ is reasonable and works well in practice. Such is the case in thermodynamics, where the origins of the method can be traced, and more recently in some applications such as collective behavior in neural populations (Schneidman et al., 2006), where new experimental and technical advances provide the means for the collection of massive amounts of data. However, in other applications the data set is so small that value space of constrains remains under sampled. Such is the case of large biological networks where sample collection is severely limited.

The surprisingly accurate descriptions of experiments provided by reformulations of some biological networks in terms of an inverse problem of statistical mechanics invites the careless application of maximum entropy methods disregarding issues arising from small sample size. The subject has driven the recent attention in the field (Haimovici & Marsili, 2015; Macke, Murray, Latham, Computational, & Unit, 2013; Nemenman & Bialek, 2004; Panzeri, Senatore, Montemurro, & Petersen, 2007).

Here we address the modeling of systems composed of a large number of interacting units from information in experimental datasets. We mainly focus on sampling bias in maximum entropy approaches with finite data sets without forcedly equating expectation values to corresponding experimental average values, and without departing from an explicit probabilistic exposition. Even though we formulate the general framework of our method, unfortunately the equations are well complicated. We confine the examples to specific simple cases, hopping clear but sufficient illustration of the concepts, unencumbered by sophisticated numerical methods.

## Theory and Methods

**Predictive distribution**: The principle of maximum entropy is connected to experiments by searching probability distributions with as little structure as possible, consistent with certain average behaviors of the system observed in the data. However, when the sample size is small, the approximation $M_j = \overline{f_j}$ is not reliable. The best we can do is to account our uncertainty about $M$. The rules of probability allows that by averaging out through marginalization with the probability of the expectation values estimated from the data $P(M|D)$. The predictive distribution can be stated as follow:

$$P(\boldsymbol{\sigma}|D) = \int P(\boldsymbol{\sigma}|M)\, P(M|D)\, \mathrm{d}M$$

By Bayes' theorem:

$$P(M|D) = \frac{P(M)\, P(D|M)}{P(D)} = \frac{P(M) \prod_i P(\boldsymbol{\sigma}_i|M)}{P(D)}$$

Given $M$, the principle of maximum entropy uniquely determines a set of Lagrange multipliers $\boldsymbol{\lambda}$ that define the probability distribution $P(\boldsymbol{\sigma}|M)$. By changing the metric

$$P(\boldsymbol{\sigma}|M) = P(\boldsymbol{\sigma}|\boldsymbol{\lambda}), \quad P(M)\mathrm{d}M = P(\boldsymbol{\lambda})\mathrm{d}\boldsymbol{\lambda}$$

Now the family of probability distribution $P(\boldsymbol{\sigma}|\boldsymbol{\lambda})$ define a class of normalizable sampling distribution parameterized by the Lagrange multipliers $\boldsymbol{\lambda} = \{\lambda_j\}$:

$$P(\boldsymbol{\sigma}|\boldsymbol{\lambda}) = \frac{e^{-E(\boldsymbol{\sigma}|\boldsymbol{\lambda})}}{Z_\lambda}$$

And

$$P(D|\lambda) = \prod_i P(\sigma_i|\lambda) = \prod_i \frac{1}{Z_\lambda} \exp\{-E(\sigma_i|\lambda)\} = \frac{1}{Z_\lambda{}^n} \exp\left\{-\sum_j \lambda_j n\overline{f_j}\right\}$$

In the statement of the problem we are supposed to know the data and nothing else, hence $\Pr(M)$ or $\Pr(\lambda)$ has to be as uninformative as possible in their respective metric. It seems to us easier to assign the prior probability $\Pr(\lambda)$. We have no preference for negative, positive or null coupling parameters, hence we assign a uniform prior probability to $\Pr(\lambda)$. Therefore, the predictive distribution can be restated as follow

$$P(\sigma|D) = K^{-1} \int P(\sigma|\lambda)\,P(D|\lambda)\,d\lambda = K^{-1} \int \frac{1}{Z_\lambda{}^{n+1}} \exp\left\{-\sum_j \lambda_j(f(\sigma) + n\overline{f_j})\right\} d\lambda$$

Where $K = \int \Pr(D|\lambda)\,d\lambda$. Hence, given the model functions $\boldsymbol{f}$ and the parameters $\boldsymbol{\lambda}$, our uncertain is exactly codified in the probability distribution $\Pr(\sigma|\overline{\boldsymbol{f}}, n)$, being $\overline{\boldsymbol{f}}$ and $n$ join sufficient statistics in the sense that further structure of the data is irrelevant here.

**Inverse problem**: It is commonly the case that our main purpose is to estimate the Lagrange multipliers. The inverse problems of inferring these parameters from the expectation values is in general hard, and are usually approached by maximum likelihood procedures.

We restate the inverse problem in terms of the posterior probability of $\lambda$ given the data with the same uniform prior probability $\Pr(\lambda)$:

$$P(\lambda|D) = K^{-1}\,P(D|\lambda) = K^{-1} \frac{1}{Z_\lambda{}^n} \exp\left\{-\sum_j \lambda_j n\overline{f_j}\right\} \tag{1}$$

**Entropy**: Our predictive uncertain can be quantified by the entropy:

$$-\sum_\sigma P(\sigma|D) \log \Pr(\sigma|D)$$

**Study cases**: We bring to discussion two toy examples, coin tossing and dice rolling, and choose a very simple Ising model, carrying the minimal required structure to illustrate the points and suggest generalization. In those cases $\boldsymbol{\sigma}$ is an array of finite discrete variables $\sigma_1, \ldots, \sigma_L$, where the quantity $\sigma_l$ can take on a discrete set of states $\{a_1, \ldots, a_r\}$, with $r \geq 2$.

# Results and discussion

The practice of using sample average as surrogates of probability expectations is reliable provided sample size is large. However, important biases can arise for small finite samples, and the issue have been the focus of attention in the specialized literature, mainly from the information theory perspective (Macke et al., 2013; Marsili, Mastromatteo, & Roudi, 2013; Nemenman & Bialek, 2004; Panzeri et al., 2007).

However, the performance of maximum entropy with surrogate sample averages is not a pathological behavior of the principle. Equating sample average to expectation implicitly changes the question we are intending to ask, and maximum entropy is given the right answer to that later question. Sample average are identical to the expectation value of a random draw from the sample itself. In the case of coin tosses, instead of asking for the probability distribution of the next toss, we can be asking "what is the probability distribution of a random draw from the thus far sample, when only some of its exact moments $\langle f_j \rangle_{sample}$ have been revealed to us?" With this unintended but different question the entire sampling space was replaced by the sample itself (over fitting), and in this case $\langle f_j \rangle_{sample} = \bar{f}_j$.

Entropy subject to restrictions on sample moments is not what we ultimately want to maximize. But it is not sense to search for a correction to that entropy trying to codify our uncertain as if expectation were actually known, since if we have that much information, it must be incorporated in the form of contrains. The entropy we can hopefully approach must account for that piece of uncertain, we are otherwise introducing more or different information than we have in our entropy constrains.

We compare the performance of maximum entropy applied to constrain of actual expectations (ME) and the empirical maximum entropy (EME) applied to sample average constrains with three simple study cases: the classical coin tossing, the dice rolling, and last a simple Ising model.

**Coin tossing**: From an experiment with a coin, we are told only the number $n$ of tosses, the two possible outcomes $\sigma = 0,1$, and the average $\bar{\sigma} = 1/n \sum_i \sigma_i$. Given the first moment $M_1 = \mu$ the predictive probability distribution, in this case the probability of the next toss being $\sigma$, is the exponential:

$$\Pr(\sigma|\mu) = \frac{1}{Z} e^{-\lambda \sigma}$$

with partition function $Z = 1 + e^{-\lambda}$. Using the substitution $x = e^{-\lambda}$ yield

$$\frac{\partial F}{\partial \lambda} = \frac{e^{-\lambda}}{1 + e^{-\lambda}} = \frac{x}{1 + x} = \mu$$

Then

$$\Pr(\sigma|\mu) = \mu^\sigma (1-\mu)^{1-\sigma}$$

The posterior probability distribution of $\mu$ with uniform prior probability, given the sample data is the Beta distribution

$$\Pr(\mu|D) = \frac{(n+1)!}{(n\bar{\sigma})! \, (n(1-\bar{\sigma}))!} \mu^{n\bar{\sigma}} (1-\mu)^{n(1-\bar{\sigma})}$$

The predictive distribution given the sample is

$$P(\sigma = 1|D) = \int \Pr(\sigma|\mu) \, P(\mu|D) \, d\mu = \frac{n\bar{\sigma} + 1}{n + 2}$$

The predictive distribution $\Pr(\sigma|D) = \Pr(\sigma|\bar{\sigma}, n)$ according to ME is a Bernoulli trial with success probability $p = \frac{n\bar{\sigma}+1}{n+2}$, which turn to be the Laplace rule of succession, with variance $var(n) = p(1-p)$. When $n$ increase the variance approach to $\bar{\sigma}(1-\bar{\sigma})$.

Performance of ME and EME are compared in Table 1 with two extreme and suggestive examples $\bar{\sigma} = 1/2$ and $\bar{\sigma} = 1$. Only ME depends on sample size n. Figure 1 plot $\Pr(\sigma|\bar{\sigma}, n)$ vs. $\bar{\sigma}$ for sample sizes $0, 1, 5, 10$ and $100$.

ME and EME perform identical when $\bar{\sigma} = 0.5$ for all sample size $n$. When $\bar{\sigma} = 1$ (or $\bar{\sigma} = 0$) EME give extreme probabilities $\delta_{\sigma,\bar{\sigma}}$. According to this EME performance, just after the first toss we are justified to attach in the next toss all the probabilities to that same "primogenital" outcome, which is obviously wrong, given the statement of the problem. After the first outcome our knowledge does not update so drastically at least. Our prior knowledge can anticipate that average $\bar{\sigma}$ will be 0 or 1 just after the first outcome, even knowing the success probability $p$, close or not to 0.5. Indeed, this radical assignment of EME does not allow for future updating by the rules of probabilities, since $\Pr(\sigma_2, \ldots, \sigma_{n-1}|\sigma_1, \sigma_n)$ becomes indefinite when $\sigma_1 \neq \sigma_n$. But even if there exist a way it can be defined,

$$\Pr(\sigma_n|\sigma_1, \sigma_2, \ldots, \sigma_{n-1})$$
$$= \frac{\Pr(\sigma_n|\sigma_1) \Pr(\sigma_2, \ldots, \sigma_{n-1}|\sigma_1, \sigma_n)}{\Pr(\sigma_n = 0|\sigma_1) \Pr(\sigma_2, \ldots, \sigma_{n-1}|\sigma_1, \sigma_n = 0) + \Pr(\sigma_n = 1|\sigma_1) \Pr(\sigma_2, \ldots, \sigma_{n-1}|\sigma_1, \sigma_n = 1)} \quad (2)$$
$$= \frac{\delta_{\sigma_n,\sigma_1} \Pr(\sigma_2, \ldots, \sigma_{n-1}|\sigma_1, \sigma_n)}{\Pr(\sigma_2, \ldots, \sigma_{n-1}|\sigma_1)} = \delta_{\sigma_n,\sigma_1},$$

even if $10^{20}$ subsequent tosses produce the same but opposite outcome in row than the primogenital one. But if we change the question as suggested in the previous section by "what is the probability distribution of a random draw from the thus far sample…?" the performance of EME in, Table 1 and the $1 + 10^{20}$ trials case fit much better to our sense. In particular the question change at each trial, since the sample space (the sample thus far from where will draw) update as $n$ increase.

On the other hand, when $\bar{\sigma} = 1$ (or $\bar{\sigma} = 0$) ME start with probability 0.5 of success, and increase (decrease) gradually to 1 (0) with sample size $n$. Our inference seems to us more affine with this performance, since we require various tosses before we seriously become suspecting the coin.

*Table 1: Predictive probability distribution obtained by ME and EME.*

| $\bar{\sigma}$ | ME $\Pr(\sigma = 1|\bar{\sigma}, n) = \frac{n\bar{\sigma}+1}{n+2}$ | EME $\Pr(\sigma = 1|\bar{\sigma}) = \bar{\sigma}$ |
|---|---|---|
| $\frac{1}{2}$ | $\frac{n/2 + 1}{n + 2} = \frac{1}{2}$ | $\frac{1}{2}$ |
| $1$ | $\frac{n + 1}{n + 2}$ | $\delta_{\sigma,\bar{\sigma}}$ |

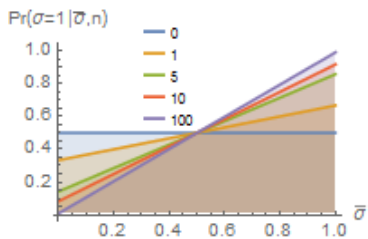

*Figure 1: Predictive probability distribution of the next toss for sample size from 0 to 100. $\delta_{x,y}$ is the Kronecker delta. The intermediate average that can't be accomplished for small sample sizes, are anyway interpolated by the same formula and and plotted.*

**Rolling dice**: Suppose we are told only a die has been tossed $n$ times and that the average number of spots up was $\bar{\sigma}$. We denote $i = \sigma - 1$ of spot up. From the partition function

$$Z = \sum_i e^{-\lambda i} = \frac{1 - e^{-6\lambda}}{1 - e^{-\lambda}}$$

$$\Pr(\sigma|\mu) = \frac{1 - e^{-\lambda}}{1 - e^{-6\lambda}} e^{-\lambda(\sigma-1)}$$

$$\Pr(\lambda|\bar{\sigma}, n) = K^{-1} \frac{1 - e^{-\lambda}}{1 - e^{-6\lambda}} e^{-\lambda n(\bar{\sigma}-1)}$$

Where $K = \frac{1}{6}(\text{Beta}[e^{6\lambda}, \frac{5-n\bar{\sigma}}{6}, 0] - \text{Beta}[e^{6\lambda}, 1 - \frac{n\bar{\sigma}}{6}, 0])$

Figure 2 shows for various combinations of data, $\bar{\sigma}$ and n, the probability distribution of the first moment $P(\mu|\bar{\sigma})$ and the probability distribution of the Lagrange multiplier $P(\lambda|\bar{\sigma})$. The uncertatinty decrease from flatish distribution ($n = 1$) to almost delta function for $n = 10000$. Even in 30 tosses showing one spot up in row ($\bar{\sigma} = 1$), some but small probability is reserved for $\langle \sigma \rangle > 1$.

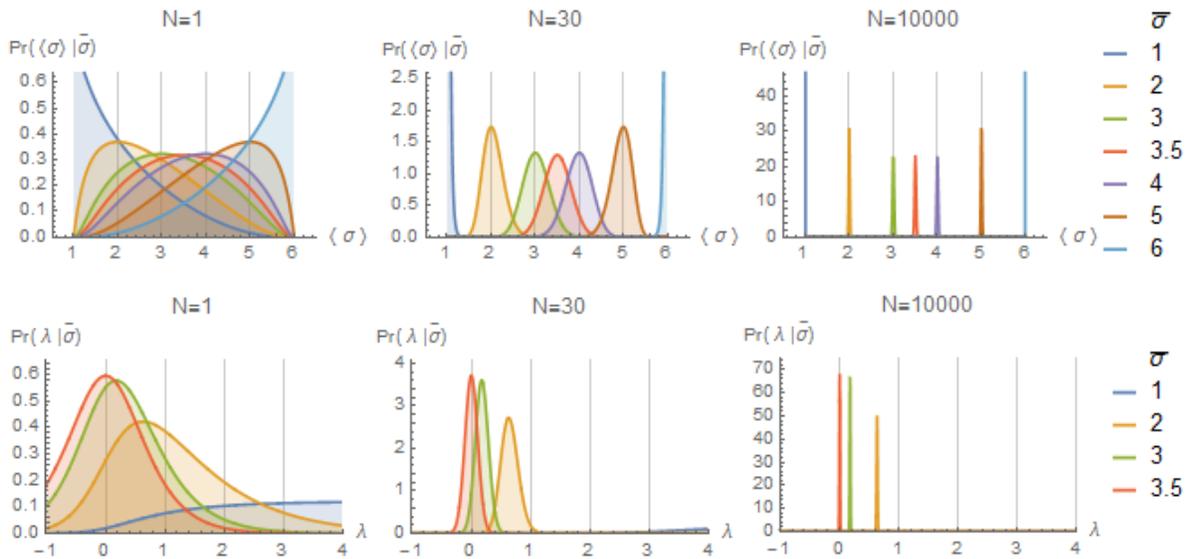

*Figure 2 Upper panel show the plots of the probability distribution of the first moment $P(\mu|\bar{\sigma})$ for various combination of data $\bar{\sigma}$ and sample size n. Lower panel show the plots of the probability distribution of the Lagrange multiplier $P(\lambda|\bar{\sigma})$ for various combination of data $\bar{\sigma}$ and sample size n.*

**Ising model**: These models can be described by an array of $L$ binary variable (spins) that are (magnetically) coupled to each other, if one spin is in the up (+1) state then its immediate neighbors could be energetically favorable to be in the same state (ferromagnetic case) or in the opposite (-1) state (antiferromagnetic case). Inverse Ising problem has received in recent years important attention, especially in inference in neuroscience, flocks, network biology, and sociology. Ising models described by the probability distribution which maximize the entropy $S$ subject to constrains such as the spin mean and the spin-pair correlation is a Gibbs distribution with energy function

$$E(\sigma; J, H) = -\frac{1}{2}\left[\sum_n H_n \sigma_n + \sum_{m,n} J_{mn} \sigma_m \sigma_n\right]$$

The Lagrange multipliers conventionally denoted $J_{mn}$ and $H_n$ correspond to the coupling of neighbors spins $\sigma_m$ and $\sigma_n$ and the mean field of spin $\sigma_n$. The predictive probability is

$$P(\boldsymbol{\sigma}|\boldsymbol{J},\boldsymbol{H}) = \frac{1}{Z}\exp\{-E(\boldsymbol{\sigma};\boldsymbol{J},\boldsymbol{H})\}$$

The inverse problem consists in estimating the values of the interaction parameters from a set of experimental Ising configurations with unknown interaction.

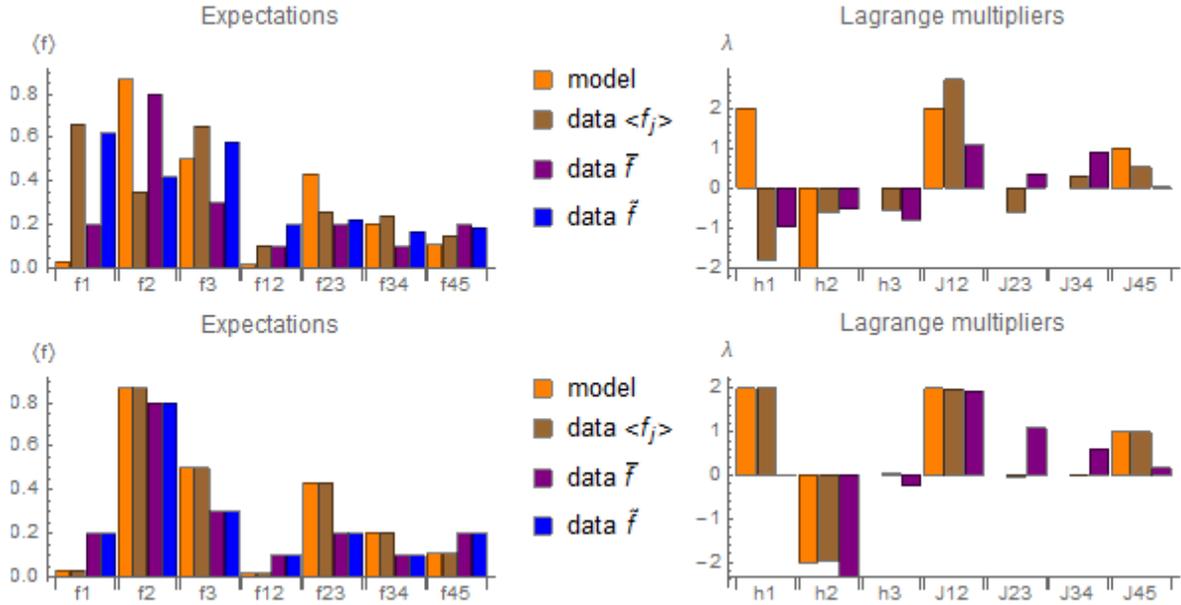

Figure 3: Prediction of the Ising model from a sample of 10 experimental configurations after 100 (upper) and 10000 (bottom) iterations. Expectation values (Left) and coupling/interaction parameters (Right) at each iterations stage. Orange and purple bar series are the actual model expectation and actual sample average values. Brown and blue series are the expectation values and sample averages as predicted from the running model after 100 and 10000 iterations.

A sample of size 10 was drawn from a simple Ising system with parameters $h_1 = 2, h_2 = -2, h_3 = 0, J_{12} = 2, J_{23} = 0, J_{34} = 0, J_{45} = 1$. Except for a slight variation, we implemented the algorithm proposed by (Mora, Walczak, Bialek, & Callan, 2010) for inferring the model parameters. The updating of the parameters at each iteration was identical, but they used Metropolis algorithm for computing the expected values from the full predictive probability distribution, while we use exhaustive computation since our smaller sampling space so permitted. Since the problem of sampling bias in maximum entropy application is not a numerical but an under sampling issue, we only need to ensure convergence. To test for convergence we compare actual expected values and sample averages with those predicted from the running model along the iterations. As shown in Figure 3 left panel, the sample averages (purple series) markedly differ from the model expectation values (orange series), as can be expected for such small sample size. Convergence is achieved after 10000 iterations as can be seen for the closeness of respective expectations and averages series (Figure 3 lower left). Model actual and predicted parameters from constrictions on expected values are very close (orange and brown series in Figure 3 lower right). However, the models parameters predicted from constrictions on sample average (purple series) are markedly different from the actual model parameters.

A sample of size 500 was randomly drawn from an Ising model with parameters $h_1 = 0.14$, $h_2 = 0.05$, $h_3 = 0.8$ and $J_{12} = 2$. The marginal joint probabilities $P(h_1, J_{12}|D)$ of the parameters $h_1$ and $J_{12}$ given the data is plotted in Figure 4. The marginal probabilities conditional on the actual model values of $h_1$ and $J_{12}$, $P(h_1|J_{12} = 2, D)$ and $P(J_{12}|h_1 = 0.14, D)$, are delineated in blue and green respectively. In particular the maximum likelihood prediction of $h_1$, crossing the top of the 2D peak, appears deviated from the value of the actual model parameter. However, these conditional distributions remain within the main support of the $h_1, J_{12}$ distribution, showing the allowance for uncertainty due to under sampling carried by equation ( 1 ).

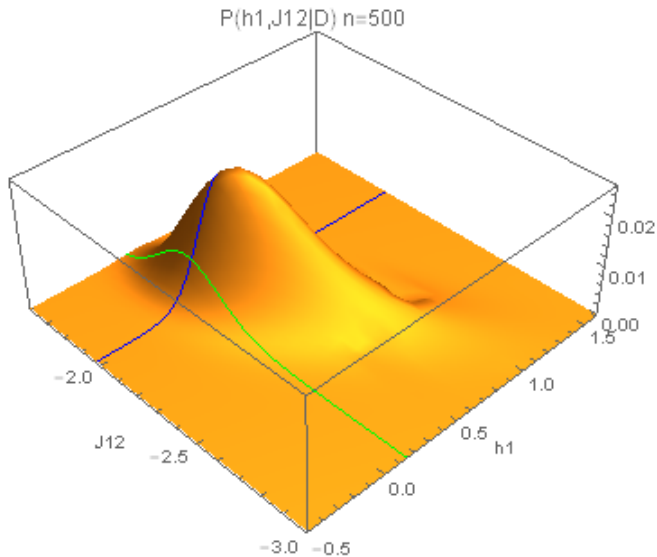

Figure 4: Marginal join probabilities $P(h_1, J_{12} | D)$ of the parameters $h_1$ and $J_{12}$ given the data. The conditional probabilities $P(h_1|J_{12} = 2, D)$ and $P(J_{12}|h_1 = 0.14, D)$ are plotted in blue and green respectively.

## Conclusions

Derived from the basic principles of probability theory, we have built a generalization of the method of maximum entropy for applications where the datasets are small. The theoretically derived model overcomes the sampling bias inherent to using sample average as surrogates of expected values in maximum entropy application.

## Acknowledgments

This work has been partially supported by grants of CIGB.